\documentclass[prl]{revtex4}%
\pdfoutput=1

\usepackage{epsfig}
\usepackage{graphics}

\begin{document}

\title{ Spacetime Indeterminacy and Holographic Noise}

\author{Craig J. Hogan}
\affiliation{University of Washington, Departments of Physics and Astronomy, Seattle WA 98195}

\begin{abstract}
A new kind of quantum indeterminacy  of   transverse position is shown  to arise from quantum degrees of freedom of spacetime, based on the assumption that classical trajectories can be defined no better than the diffraction limit of Planck scale waves. 
 Indeterminacy  of the angular orientation of particle trajectories due to wave/particle duality at the  Planck scale leads to indeterminacy of a nearly-flat spacetime metric, described as a small nonvanishing quantum commutation relation between transverse position operators at different events along a null trajectory.    An  independent derivation of the same effect is presented based on the requirement of   unitarity in black hole evaporation. The indeterminacy  is interpreted as a  universal   holographic quantum spacetime noise, with a frequency-independent spectrum of  metric perturbation amplitude, $\langle h_H^2\rangle^{1/2} \simeq \sqrt{l_P}=2.3 \times 10^{-22} /\sqrt{\rm Hz}$,  where $l_P$ denotes the Planck length. The effect  is  estimated to be directly measurable  using current interferometer technology similar to LIGO and LISA.
\end{abstract}
\maketitle
\section{Introduction}

Although string theory is sometimes referred to as a fully quantum theory of gravitation, the classical limit of the theory, in particular the emergence\cite{Seiberg:2006wf,Padmanabhan:2007tm}
 of an approximately classical 3+1 dimensional  spacetime  from the collective behavior of an essentially quantum system, is not at all understood.  This understanding would be considerable advanced by obtaining  concrete data about quantum deviations from classical behavior of  spacetime  that  directly expose its quantum degrees of freedom.
This paper offers two derivations predicting a new quantum behavior of  nearly-flat spacetime, an indeterminacy of angle or transverse position \cite{Hogan:2007rz,Hogan:2007ci}: one  based on wave mechanics at the Planck scale, the other   based on no information loss during black hole evaporation.   It then shows that the effect can be described as a new kind of quantum noise, and  analyzes how its effects might be measured using current technology.

Quantum mechanics based on continuous classical spacetime works to the limit of the highest energies and shortest distances yet  tested.   In  conventional quantum field theory, classical spacetime  is predicted to be a good approximation down to the Planck length $l_P=\sqrt{\hbar G_N/c^3}= 1.616\times 10^{-33} {\rm cm}  $, the wavelength where a single quantum has about the size and energy  of a black hole.  In particular perturbative field theory suggests that  positions in space can be locally defined and  measured to Planck scale precision.

However,  there are theoretical  reasons to suspect a larger departure from classical spacetime behavior when large regions of space are considered.
The fundamental quantum description of the world is a ray in a Hilbert space, obeying unitary evolution.   This space has a very large dimensionality, say  $ 2^N$, where $N$ is the number of binary degrees of freedom, and  it is not known how those degrees of freedom map onto a quantum spacetime that   approximately resembles a classical 3+1 dimensional manifold containing quantized fields. Arguments  from black hole and string physics\cite{Bekenstein:1972tm,Bardeen:gs, Bekenstein:1973ur,Bekenstein:1974ax,Hawking:1975sw,'tHooft:1985re,Strominger:1996sh,'tHooft:1993gx,Susskind:1994vu,'tHooft:1999bw,Bigatti:1999dp,Bousso:2002ju,Maldacena:1997re,Witten:1998zw,Aharony:1999ti,Alishahiha:2005dj,Horowitz:2006ct,Padmanabhan:2006fn}, as well as boundary behavior of classical general relativity\cite{Padmanabhan:2007en,Padmanabhan:2007xy} and thermodynamics in nearly flat spacetime\cite{Jacobson:1995ab},  suggest that the 
maximum entropy $N$ of a finite system, corresponding to its total true number of degrees of freedom,   scales holographically\cite{Bousso:2002ju}: for example,  a   region of spacetime bounded by null surfaces has total entropy less than one quarter of the area of  a bounding  2D surface in Planck units,  much less than  would be the case for the entropy of quantized fields. The number of degrees of freedom of a large ($L>>l_P$) holographic spacetime is far smaller than it would be in a field theory extrapolated to the Planck scale.  

 This information deficit provides an important clue to observable    consequences since the dimension of the Hilbert space is directly related to the number of observables. It is an open question how this quantum behavior of large spacetime volumes manifests itself to observations from within the spacetime.  Up to now, there have not been concrete predicted observable consequences of the holographic entropy bounds, or indeed of any other quantum property of spacetime.  
 
 The holographic bounds require  new correlations between observables.   The new correlations may introduce   clearly defined and predictable observational signatures, and perhaps may even be practical to measure in real experiments.   One  resulting observable quantum behavior of spacetime is a {\em Holographic Uncertainty Principle} \cite{Hogan:2007rz}, an indeterminacy in transverse position.   The  effects of spacetime quantization on the Planck scale   in some macroscopic experiments appear on  scales much  larger than the Planck length--- large enough to  be measured.  
This paper derives   the effect using general  principles of wave optics; formulates it in terms of a quantum commutation relation between transverse position observables in free space; and  sharpens its interpretation and predicted experimental consequences.  In  particular, it is predicted that spacetime is permeated with  a universal ``holographic noise'' accessible to measurement by realistic experiments.

The discussion below begins  with holographic indeterminacy   estimated from diffraction theory of particle trajectories emerging  from Planck-wavelength radiation.  Quantum mechanics is built on the idea of   wave-particle duality,  but classical spacetime  is a manifold defined by classical paths and definite trajectories connecting separated points. In spacetime defined by propagation of quantum waves,
``blurring'' in angular eigenstates arises  from simple particle/wave optics:  it is not possible to measure the orientation of a  trajectory of a particle over a distance $L$ to a precision better than $\Delta \theta\simeq (l_P/L)^{1/2}$. There is thus a quantum limit to the precision to which  classical geodesics can be defined. 
This picture shows the transverse character of the resulting indeterminacy, and demonstrates the key idea advanced here:  Planck-scale wave effects are optically magnified to much larger scales in experiments with macroscopic baselines.  

Observable indeterminacy is then more rigorously analyzed   with a simple quantum algebra that introduces a new nonvanishing commutator between operators for the     transverse position observables.  In the usual way this leads to complementarity and uncertainty relations between these observables. An unusual feature of the interpretation is that these relations hold for any body or particle, since they represent quantum indeterminacy of spacetime itself.   The wavefunctions of spacetime states can be  measured by experiments, and this formulation allows statistical predictions for experimental outcomes.

An independent derivation of the effect is then presented,    based directly  on the assumption that  the black hole evaporation process conserves total degrees of freedom.    In order for any black hole and its evaporation products together to obey unitary evolution, the angular position of evaporation products in flat space must display quantum indeterminacy: otherwise there would be more distinguishable ways to assemble a hole than there are black hole states. Holographic uncertainty in angle or transverse position then follows from the time reversal of these continuous trajectories with an observer put in place of the black hole. This argument allows a quantitative estimate  of the  numerical coefficient of the uncertainty.

Finally, estimates are presented of  holographic signals in interferometers that make nonlocal comparisons of the transverse   positions of widely separated bodies.  Spacetime indeterminacy is shown to create a pervasive universal holographic quantum noise in the metric with a   power spectral density given by the remarkable formula $\langle h_H^2\rangle\simeq l_P$.    Estimates  of  the response of the Laser Interferometer Space Antenna  ({ LISA}) suggest that holographic noise can be separated from  LISA's instrumental, environmental and gravitational noise well enough to be measured.  It may also be possible to develop purpose-built ground based experiments capable of measuring the effect on a smaller scale, so that the spectrum and spatial character of the noise can be characterized.    
 If the effect is detected experimentally, it will confirm the  holographic character of unification and allow detailed, quantitative experimental studies of quantum gravity. 
\section{Angular Indeterminacy from Planck Diffraction Limit}

Quantum gravity is sometimes described as a system of Planck scale waves. Local measurements in such a system can clearly be made with resolution  to the Planck length, consistent with the Heisenberg uncertainty principle.  However even in this simple system,  nonlocal positions cannot be compared with Planck length precision. They are subject to uncertainty on  a larger scale due to diffraction. 

The effect can be described in terms of the emergence of particle trajectories from wave optics.  A classical spacetime is a manifold with well defined paths such as geodesics; indeed, its metric properties are defined by paths composed of summed infinitesimals.  If at a fundamental level spacetime is defined by interactions and propagation of Planck scale quantum waves, the classical paths emerge as rays that have a fundamental indeterminacy in their angular orientation, due to their wave properties. This leads to  a corresponding indeterminacy in transverse position.

\begin{figure}
\epsfysize=1in 
\epsfbox{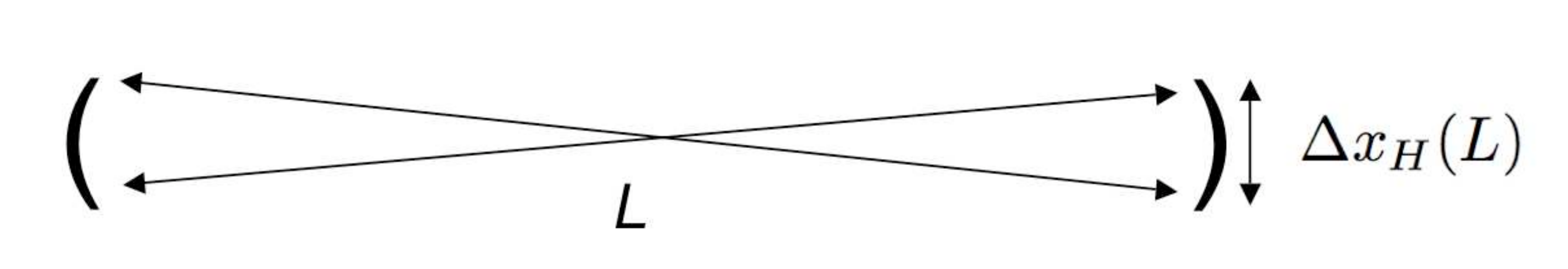} 
\caption{ \label{diffraction}The holographic indeterminacy in transverse position $\Delta x_H(L)\simeq(Ll_P)^{1/2}$ is related to the   diffraction limit of Planck wavelength radiation. Wavefronts with radius of curvature $\simeq L$ are curved by $\simeq l_P$ across an aperture of size $\Delta x_H(L)$.   Universal indeterminacy of the angular orientation of a particle trajectory  arises because smaller apertures are limited by diffraction and larger apertures are limited by quantum indeterminacy  of particle position within an aperture; thus over length $L$ the angular orientation of any particle's trajectory is intrinsically uncertain by $\Delta \theta (L) \simeq (l_P/L)^{1/2}$.  An aperture of this transverse size  creates a diffraction limited spot of Planck radiation equal to its own size at distance $L$.}
\end{figure}

Consider the propagation and imaging of  waves of wavelength $l_P$.
A Planck length (or equivalently, a Planck time) of phase delay gradient  or tilt added to a plane wave across an aperture of size
\begin{equation}
\Delta x (L)= (L l_P)^{1/2}
\end{equation}
 changes the normal direction  by the angle
 \begin{equation}
\Delta \theta (L) = (l_P/L)^{1/2}.
\end{equation}
This angle and transverse length define the {\em Planck Diffraction Limit} for a distance $L$: an aperture of size $\Delta x (L) $ creates a spot of minimum transverse size $\simeq\Delta x (L)$ at distance $L$, and can make images that resolve angles as small as $\Delta \theta (L)$. A larger aperture can resolve a smaller angle; on the other hand due to quantum indeterminacy,   with a larger aperture one cannot tell where in the aperture any particular particle arrives. Due to quantum complementarity, if one tries to determine where in an aperture a particle path lies, the angular precision deteriorates to that of a smaller aperture. (This is the same consequence of wave-particle duality that is familiar from the double slit experiment: you cannot tell which slit a particle passes through and still preserve the interference pattern from the slits.)  Thus {\em no   measurement can define  the angular orientation of  a single particle trajectory over length $L$ better than $\Delta \theta (L) $, leading to a fundamental indeterminacy  in defining classical paths over a length $L$.}  Correspondingly,  $\Delta x(L)$ is the smallest transverse distance where two observers separated by distance $L$ can resolve each other with Planck wavelength radiation, and the smallest transverse position difference that can be  classically defined at that separation (See Fig.  \ref{diffraction}).  

Because of the angular indeterminacy, the classical spatial direction of a particle propagating from an event only becomes clear over time as the angular uncertainty gradually decreases.   This is a simple example of ``emergence'' of a classical path from a wave description.   
The complementary effect advanced here is  branching of spacetime metrics:  transverse spatial positions  decohere from each other over large spacetime separations. 

This simple model   illustrates how wave effects at the Planck scale can be optically magnified to much larger scales over macroscopic distances $L$. It displays the  magnitude of the effect as well as the transverse character of the uncertainty.
Thus we can  visualize holographic uncertainty  as an effect caused by diffraction of  Planck scale waves out of which a classical spacetime emerges.     It is conjectured that since it affects all the paths that define a classical spacetime,  all bodies  acquire  a  quantum interderminacy in transverse position.

\section{Noncommuting  Spatial Position Observables}

\begin{figure} 
\epsfysize=3.5in 
\epsfbox{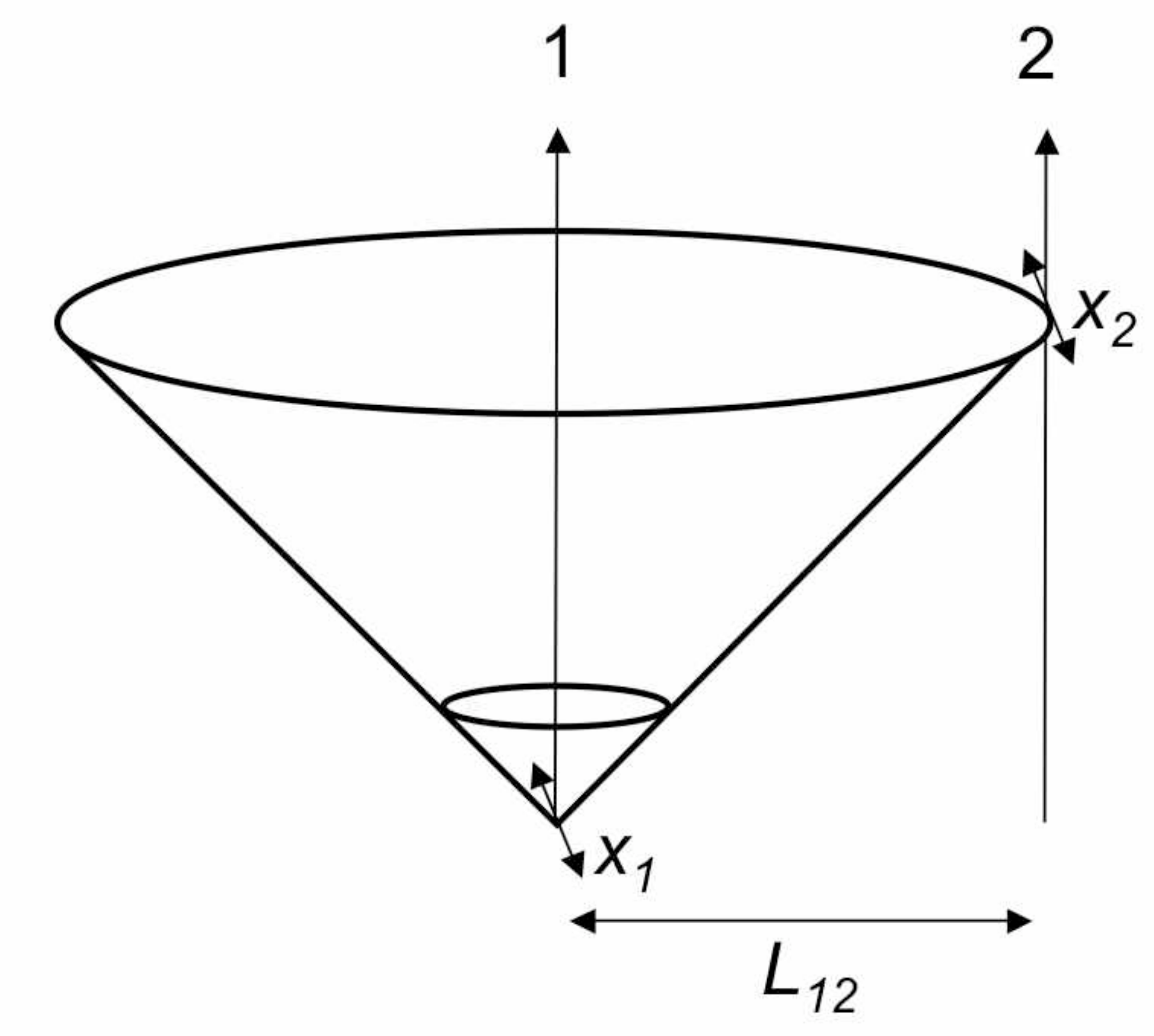} 
\caption{\label{cone}
Coordinates used in Eq. (\ref{commute}).   World lines in spacetime are shown for particles 1 and 2; commutation relation refers to transverse position operators at events on a light cone. In this figure the $y$ transverse dimension is suppressed.}
\end{figure}

It is useful to construct a simple quantum theory of holographic indeterminacy,  based on a nonvanishing commutation between spatial observables of position.  Consider two particles at rest relative to each other separated by distance $L_{12}$.   Introduce  transverse position operators $\hat x_1$  and $\hat x_2$ for the two particles in the same spatial direction $x$ orthogonal to their separation vector. Let $\psi(x_1,x_2)$ denote a wavefunction, the amplitude for the particles to be at   positions $(x_1,x_2)$. We    interpret $\psi$ not only as a particle wavefunction, but as a spatial wavefunction of any possible particles--- the amplitude that   particles or a bodies   at  separation $L_{12}$ lie at transverse positions $x_1,x_2$  relative to their classical separation vector.
 In normal quantum mechanics in unquantized spacetime these are commuting observables,
$[\hat x_1,\hat x_2]=0$; a classical spacetime by its nature has well defined positions so its wavefunction is an eigenstate of both, $\hat x_1 \psi(x_1,x_2)=x_1\psi$ and $\hat x_2\psi(x_1,x_2)=x_2\psi$. 
 (Indeed, classically, the particles by definition have vanishing transverse displacement from their separation vector so $x_1=x_2=0$; in normal quantum mechanics on a classical spacetime background, depending on the  state of particles there is generally uncertainty in position so the displacement observables no longer vanish, but their commutator still does.)  Posit now that spacetime quantization results in  a new small but nonvanishing commutator between the position operators,
\begin{equation}\label{commute}
[\hat x_1,\hat x_2]=-i l_P L_{12}.
\end{equation}
To be covariant, this relation is understood to refer to the observables evaluated at times corresponding to events separated by a null trajectory (see Figure \ref{cone}). (Note that this commutator differs from others recently proposed to  mimic holographic behavior\cite{Jackson:2005ue,Jackson:2007tn}.)

It can be seen that this relationship is simply a more precise statement of  the indeterminacy analyzed in the previous section.  We again suppose that classical spacetime is an emergent system and that classical paths are defined by rays of Planck wavelength massless particles with  energy $p_0=l_P^{-1}$. Flat spacetime is defined by straight trajectories so up to a  phase, the spatial quantum wavefunction  for a family of parallel, classical spacetime null paths  corresponds to a plane wave,
\begin{equation}
\psi = \exp[ i ({ p_0t-\bf px})],
\end{equation}
where $\bf x$ denotes 3-position  and  $\bf p$ denotes the 3-momentum vector. 
A particular classical path or particle trajectory connecting events  is specified by a definite 3-direction $\bf\hat p$ and by a transverse position.
Label the three spatial directions $x,y,z$, and let $\bf\hat p$ lie along the $z$ axis, so transverse positions lie  in the $x,y$ plane. Without loss of generality, choose a classical path with $x=y=0$ and zero transverse momentum, $p_x=p_y=0$. Spacetime wavefunctions describe deviations from   classical paths by having a nonzero amplitude for   transverse momentum and transverse spatial displacements.
At any time $t_1$  a Planck particle has the usual Heisenberg commutation relation between transverse momentum and position in the same direction,
\begin{equation}\label{heisenberg}
[x_1,p_{x 1}]=-i.
\end{equation}
 The transverse momentum  $p_{x 1}$ of a Planck particle at event 1 is related to the transverse position displacement at event 2 by the angular deflection of the plane wave,
\begin{equation}\label{deflection}
 p_{x 1}/p_0= p_{x 1}l_P= x_2/ L_{12},
\end{equation}
and combining equations (\ref{heisenberg}) and (\ref{deflection}) 
at any two positions $t_1, x_1$ and $t_2, x_2$ on a light cone yields Eq. (\ref{commute}). Again the ``particle'' in this exercise is a proxy for a quantum behavior of paths in  the spacetime metric, rather than an observable physical particle. The relation (\ref{commute}) is posited to hold for any position measurements.    

Clearly these simple arguments do not constitute a fundamental theory of emergent spacetime.  Nevertheless, they are sufficient to calculate  new phenomenological consequences of holographic spacetime quantization.  Equation (\ref{commute}) is  conjectured to apply because it  obeys required symmetries, matches holographic bounds on degrees of freedom, and (as explained below) is consistent with local and global behavior of quantum states during black hole evaporation.   The exact numerical factor is not determined by general considerations, although a lower bound of order unity is derived 
 in the following section.
 
The relation is defined for particles at rest which defines the transformation properties in other frames.   In particular a large boost along the separation vector, which shrinks $L_{12}$ in the observer frame, correctly reduces to small indeterminacy in the limit of a local experiment.   In the $x$ direction, Eq. (\ref{commute}) makes an assertion about transformation properties for bodies not at rest in the observer frame: the indeterminacy in $x$  is independent of the $x$ boost. In other words, it does not depend on the relative transverse motion of the body and the observer but is  again consistent with being an invariant property of spacetime. Moreover it is independent of particle momentum, and therefore mass, as required for an indeterminacy in position associated with all bodies in space regardless of their specific properties.  Like the diffraction indeterminacy  just discussed, it leaves the two transverse directions $x$ and $y$ independent of each other. 

The two observables $x_1,x_2$ now become complementary variables subject to quantum indeterminacy via the Heisenberg uncertainty relation
\begin{equation}\label{uncertainty}
\Delta x_1\Delta x_2>l_P L/2.
\end{equation}
Here $\Delta x_1,\Delta x_2$ denote the width of the wavefunction in the two positions, and therefore the joint quantum indeterminacy in   measurements of the position observables in a particular state.  This expresses the same indeterminacy discussed above in the context of Planck diffraction.  
A system can be put into a squeezed state where one of the observables has a small uncertainty, but the sum of the uncertainties is minimized when $\Delta x_1=\Delta x_2$ and therefore $\Delta x_1^2> l_P z/2$, yielding the holographic uncertainty relation. 
This is also   the typical state of quanta detected from evaporating   black holes,  as described below. 

Up to a numerical factor, this indeterminacy yields entropy limits consistent with  covariant entropy bounds.   For a sphere of size $L$, particle trajectories are defined only to within  solid angle $\Delta \theta^2\simeq l_p/L$, yielding $\simeq L/l_P$ independent directions for quantized modes.  Each direction has $\simeq L/l_P$ modes up to a Planck scale cutoff, so there are  $\simeq (L/l_P)^2$ independent modes  or degrees of freedom.
In normal field theory, there would be $\simeq (L/l_P)^2$ independent directions and $\simeq (L/l_P)^3$ degrees of freedom.

  Since $\psi$ is interpreted as a spacetime wavefunction, the  shape of $\psi$ is shared by all bodies in a local region to a precision given by  Eq. (\ref{commute})  where $L_{12}$ is the size of that region.  This broader interpretation is necessary because any purely local experiment cannot detect the holographic uncertainty associated with a distant measurement.  That is, $x,y$ positions commute locally so   bodies within a local region cannot have different holographic wavefunctions; they have to be squeezed into the same spacetime state   to a precision given by the holographic uncertainty for that local region.  Thus the whole spacetime metric (and all the bodies in it) shares in the new quantization reflected in Eq. (\ref{commute}) and uncertainty reflected in Eq. (\ref{uncertainty}).  
  
  Similarly, observations that ``collapse'' the  spacetime wavefunction also collapse all $x,y$ positions nonlocally into the same spacetime eigenstate,   up to a precision given by the coherence scale $L_{12}$.
For a covariant formulation, the  radial dependence should be interpreted as a spreading of $\psi$  with separation,  a ``many-worlds'' quantum branching  of metrics with propagation along a light cone.  This interpretation preserves the agreement of the effect with covariant entropy bounds, which are defined in terms of light-cone volume.  This property has consequences for the spatial and temporal coherence scales of the observable  uncertainty, as described below:
measured displacements $\Delta x(L)$ are coherent over a temporal and spatial extent of order $L$.

In this formulation of the holographic indeterminacy, transverse position, which is normally a classical observable that commutes with   position  operators at different points, now exhibits noncommuting quantum behavior. 
In this interpretation   there is no ``true'' classical transverse position that is merely blurred by propagation;   any transverse position is quantum-indeterminate without a true classical value. 
The difference is important in experimental design since a true quantum indeterminacy in position can be measured interferometrically.  If the   transverse sensitivity is precise enough, we can measure quantum-gravitational  commutation relations applied to transverse position observables of macroscopic bodies.

\section{Holographic indeterminacy from unitarity of  black hole evaporation}

The simple behavior conjectured in Eq. (\ref{commute}), and its interpretation as a spacetime indeterminacy, are supported by
consideration of what happens during black hole evaporation. Although in principle the holographic information deficit could appear in subtle and unobservable nonlocal correlations between particle states,  the trajectories (and states) of evaporated particles  from black holes evolve in a way consistent with the specific and simple behavior suggested by holographic indeterminacy  in transverse position or angle.

The number of distinguishable   angles at large distance can be related to counting the quantum states of the particles  evaporated from a black hole. One can visualize independent ``pixels'' on the surface of a hole, each of  area $ l_P^2 4\ln 2$;  black hole thermodynamics tells us that the number of such pixels is the number of binary quantum degrees of freedom of the black hole, or the logarithm of the Hilbert space dimension of the quantum hole state\cite{'tHooft:1999bw}. Once such a hole has completely evaporated, unitarity of quantum evolution requires  the number of degrees of freedom of the evaporated products to equal those of the hole.

For  a black hole of radius $R_H$,   define as above a {\em Planck diffraction distance}  $L_H=   R_H^2/l_P$.  As described above, this distance is about where
one Planck wave of  wavefront curvature across an  aperture  of radius $R_H$ corresponds to a radius of curvature $L_H$.  If the black hole were a lens, with waves longer than a Planck length  it would not be able to distinquish angles smaller than $\Delta\theta_H=R_H/L_H$ due to Planck scale diffraction.

The hole's ``pixels'' each subtend solid angle $\pi \Delta\theta_H^2$.
Thus there are not enough degrees of freedom in the hole to separately specify all the distinguishable distant angular trajectories that would classically be available to particles evaporating from the hole, beyond the distance $\simeq L_H$.  Reverse the  trajectories to consider inwards moving particles: for any shell beyond $L_H$,  in a classical spacetime the number of  distinguishably different directions  from which a particle of wavelength less than or equal to $R_H$ can aim at and still hit the hole exceeds the number of pixels in the hole. Thus  {\em were there no holographic uncertainty, it would be possible to find many distinguishable inward trajectories for particles (coming from different angles separated by less than $\Delta\theta_H$, from beyond $L_H$) that would end up creating exactly indistinguishable black hole states.} Since this would violate unitarity, there must be a contradiction.  

\begin{figure} 
\epsfysize=3.5in 
\epsfbox{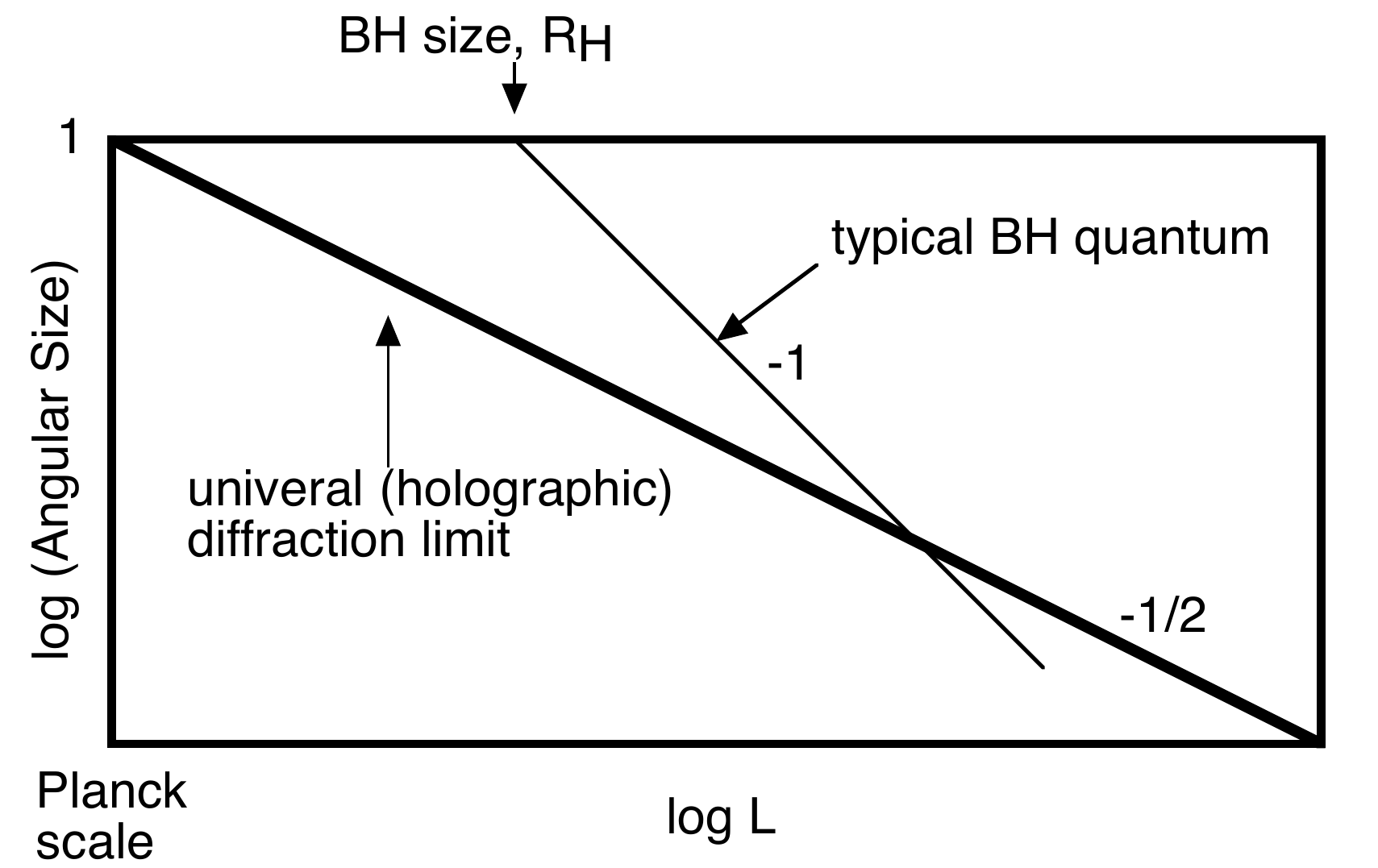} 
\caption{\label{angleplot}Angular size as a function of separation. From a  black hole,  the angular extent of a typical $R_H$-scale evaporated particle   wavefunction scales like $R_H/L$ until the fundamental  universal ``Planck diffraction limit'' is reached, and like $L^{-1/2}$ beyond that.  In this way the evaporated particles have the same entropy as the hole. Note that the universal limit also implies that classical trajectories only gradually emerge with a well defined angular orientation when they are much longer than a Planck length.}
\end{figure}

Holographic uncertainty   offers a way to resolve the contradiction in a way that respects the continuity of the trajectories of evaporated particles.  The distant spacetime does not have so many independent degrees of freedom: all  those apparently different inward traveling particle states are not actually independent states.  
The problem is solved for black holes of any size if  the holographic  limit on the number of angular states as a function of distance applies to flat spacetime generally. (see Fig.  \ref{angleplot}). As discussed above this would be a consequence of the new commutation relation of spatial position.

From the point of view of the black hole, or for that matter any observer,  it is an illusion to suppose that smaller angles at large distance have any objective classical meaning.  A typical photon from the evaporation has a wavelength $R_H$ and carries about 1 bit of information, entangled with the remaining hole.\cite{Page:1993wv}  (Longer wavelengths are suppressed by a graybody factor in the Hawking radiation formula.)  Far from the hole, there is not enough information to distinguish smaller angles than $\Delta\theta_H\simeq R_H/l_P$ for all the photons.
Internally-identical hole states produce particles having a range of transverse positions at larger distances which therefore must not actually be distinguishably different after all to the hole. 
A distant observer can measure those precise transverse particle positions, but that observer then inhabits a branch of the wavefunction that no longer includes a definite position for the hole, which is itself subject to the same transverse positional uncertainty from the point of view of the observer.
Small angular differences distinguishable far from the hole    cannot be distinguished by the black hole itself.

A more detailed version of this argument allows a quantitative estimate for the numerical coefficient for the  holographic uncertainty by counting the number of degrees of freedom of a black hole and its evaporation products.  Let the dimension of the black hole Hilbert space be $D_H=2^{N_H}$. Let the dimension of the Hilbert space of its evaporata  be $D_\gamma=2^{N_\gamma}$.  Describe the degrees of freedom $N_\gamma$ in terms of radial and transverse eigenstates, $N_\gamma=N_\| + N_\perp$.  The radial part for each photon describes its arrival time/radial position and/or its energy/radial momentum.  For unitary evolution we require $D_H=D_\gamma$; the states of the evaporata  also apply after the hole is gone so for this argument we do not need to consider ``entanglement entropy'' which mixes the hole and evaporate states \cite{Page:1993wv}.  We know the entropy of the black hole which tells us that $N_H=A/(l_P^2 4\ln 2)$, thus
$N_\perp<A/(l_P^2 4\ln 2)$, where $A=4\pi R_H^2$.  We conjecture that holographic uncertainty is responsible for the limited number of $N_\perp$ states, such that at distance $L$ the number of states is $N_\perp=4\pi L^2/\pi\Delta x_\perp (L)^2$.  Furthermore  we should only count states actually coming from the hole (those  whose time reverse trajectories hit the hole) so we set $\Delta x_\perp=R_H$.  
Thus in the end we equate  the entropy of a black hole of radius $R_H(L)$ (the of number degrees of freedom as measured by pixels in the event horizon)  to  the number of angular directions available in a sphere of  radius $L$ for a hole of this size (as measured by the ratio of the sphere area $4\pi L^2$ to black  hole cross section $\pi R^2$):
\begin{equation}
S_H(A=4\pi R_H^2)={4\pi R_H(L)^2\over l_P^2 4\ln 2}={4\pi L^2\over \pi R_H(L)^2};
\end{equation}
this  is satisfied with the universal transverse uncertainty,
\begin{equation} \label{eqn:transverse}
\Delta x_H(L)=R_H=(4\ln 2/\pi)^{1/4}(Ll_p)^{1/2},
\end{equation}
and a  width for the  angular uncertainty at distance $L$, 
\begin{equation}\label{eqn:angle}
\Delta\theta_H (L)=  \Delta x_H/L_H=  (4\ln 2/\pi)^{1/4}( l_p/L)^{1/2}.
\end{equation}
A black hole of radius $R_H$ only starts to ``notice'' the holographic uncertainty beyond a distance 
\begin{equation}
L_H= (4\ln 2/\pi)^{-1/2} (R_H^2/l_P).
\end{equation}
This estimate  yields a numerical prefactor close to unity as assumed below.
 
\section{Detection of Holographic  Noise}

\subsection{Spectrum and Character of Universal Holographic Noise}

The hypothesis of holographic indeterminacy, Eq. (\ref{commute}),  and its interpretation as a branching of metrics predicts new physical  effects that appear to be detectable with current technology.

The arguments above suggest that holographic indeterminacy should be regarded not as just a propagation effect on a classical background but as an  intrinsic quantum indeterminacy in transverse position. We should think of a family of classical metrics or ``histories''\cite{Gell-Mann:1995cu,Hartle:2002nq,Bosse:2005un} that peel away from each other as light cones expand, such that at large distances there is a superposition of   transverse positions rather than a sharply defined classical position eigenstate.  In that case it is possible to study the effect using measurements of transverse position of macroscopic bodies,  using interferometry.  Measurements of position that appear classical in small spacetime volumes   are predicted to show holographic indeterminacy over  macroscopic distances and times. 
 
Any kind of interferometer measures only differences of light paths, each of which is a radial (and not a transverse) distance. On the other hand in some configurations, the differences between these radial distances does measure the precise transverse position   of a body. If holographic uncertainty represents an intrinsic uncertainty in spacetime position states, it should be measurable as an added quantum indeterminacy in transverse position measurements.  An interferometer with arm length $L$  displays a typical transverse positional uncertainty
 \begin{equation} \label{1meter}
\Delta x_H\simeq (l_PL)^{1/2}= 4\times 10^{-16}(L/ 1{\rm m})^{1/2}{\rm \ \  cm},
\end{equation}
approximately coherent over length $L$ for a time $L$.
In  terms of metrological accuracy, (radial) positions are measured to 
this precision by interferometers currently in operation--- for example  LIGO, the Laser Interferometer Gravitational wave Observatory \cite{Abbott:2003vs,Abbott:2005ez,Abbott:2006zx}. Detectability depends both on the system noise and the configuration.

We can make the general statement that any triangle configuration that measures the relative  positions of three proof masses in a 2D plane must be accompanied by holographic  indeterminacy in those positions to agree with holographic entropy  bounds. This can be seen simply by counting spatial pixels.  At holographic  measurement precision,  the number of distinguishable positions  in  a triangle of of scale $\simeq L$ is $\simeq (L/\Delta x)^2\simeq L/l_P$, the maximum allowed for a system confined to a plane. Including the same precision in the orthogonal direction or allowing for rotations of the triangle leads to $\simeq (L/l_P)^2$  distinguishable positions in a 3D volume of scale $L$, the maximum number of degrees of freedom permitted by covariant entropy bounds.  {\em Without a holographic uncertainty in measurement precision,  a triangle interferometer would be able to measure a larger number of distinguishable  spatial positions than allowed by holographic bounds}.

Consider two beams heading towards each other across arm 1 of a triangle. They traverse an identical path in opposite directions, then are directed to the third point of the triangle along arms 2 and 3. Their travel time on those arms, which is measured when they come together,  is affected by the holographic indeterminacy of the transverse positions of the ends of   arm 1. Two beams on the same path leaving later, by more than a coherence time, will be affected by indeterminacy in the same way but with new random values.  Thus an experimental signature of holographic indeterminacy is a new source of noise in some kinds of position measurements.

 The effects  of holographic indeterminacy can be expressed as  new kind of quantum noise in the metric, with a universal spectrum.  Like gravitational radiation, holographic noise is a nonlocal effect,  not detectable in any local experiment. Similarly we can discuss it in terms of an equivalent fractional distortion of  the metric.  In an experiment on scale $L$, measurements of positions at times separated by more than $L$ are associated with the transverse uncertainty $\Delta x_H\simeq (l_PL)^{1/2}$ and hence a dimensionless  transverse  fractional distortion $h=\Delta x_H/L \simeq (l_P/L)^{1/2}$.  This can be regarded as  a result of an equivalent noise spectrum in the classical metric associated with indeterminacy in  defining classical paths and position observables.  Let $\langle h^2_H(f)\rangle$ denote the perturbation power spectral density per frequency bandwidth as a function of frequency $f$,  and let $h_{H,rms}=\langle h^2_H(f)\rangle^{1/2}$ denote    the measured root-mean-square holographic metric noise.   In a detector with scale $L$, the estimates above of the coherence and amplitude of the indeterminacy suggest that in a time $L$, the typical displacement is 
 $\Delta x_H\simeq (l_PL)^{1/2}$;  the detected spectrum $h_{rms,det}$
  is independent of frequency $f$ at low frequency,
 \begin{equation}
h_{rms,det}\simeq h_{H,rms}\simeq (l_P/L)^{1/2} L^{1/2}, \qquad (f<L^{-1}),
\end{equation}
and falls off at high frequency like
 \begin{equation}
h_{rms,det}\simeq (l_P/L)^{1/2} f^{-1/2}, \qquad (f>L^{-1}).
\end{equation}

We interpret this result as a  universal power spectrum of holographic quantum  noise in the spacetime metric itself,  inherent in the indeterminacy of transverse positions. For any system on scale $L$, at low frequencies 
(that is, for $f<L^{-1}$) the universal spectrum is flat with frequency and is given by the remarkably concise expression,
\begin{equation}
\langle h_H^2\rangle\simeq  l_P,
\end{equation}
or in more familiar experimental units,
\begin{equation} \label{noise}
h_{H,rms}\simeq \sqrt{l_P} = 2.3 \times 10^{-22} /\sqrt{\rm Hz}.
\end{equation}
It should be emphasized that apart from numerical factors of the order of unity, and geometrical factors depending on the system configuration and the spatial character of the noise, this spectrum is predicted with no free parameters. 

The detectability of this holographic noise depends on the sources of system noise and the configuration of an interferometer. It is a quantum noise, so its effects on observable signals depend on the correlations of nonlocal position observables measured by an apparatus.  The estimates   below lead to optimistic assessments of the possibility of experimental
  studies of its spectrum and spatial character.
  
 Although holographic noise is an effect of  spacetime quantization, on scales larger than the Planck length it is much larger than the effects of  virtual quanta that would be predicted from a standard quantized spin-2 graviton field. 
The spatial character of the metric disturbance is also apparently not   the normal transverse-traceless modes of classical gravitational waves ($h_+$ and $h_\times$ polarizations) or their spin-2 quantized  counterparts, but  is better described as shear modes, with no propagating classical counterparts.   

\subsection{Detectability with LIGO-like interferometers}

In the case of LIGO, the coherence time for the indeterminacy is about a light travel time over 4 km (smaller for the 2 km interferometer)  corresponding to a frequency of about 77 kHz.  At lower frequencies, the metric perturbation amplitude spectrum contributed by holographic noise thus corresponds  to the universal value, Eq. (\ref{noise}). 
  This is  a higher noise level (by about a factor of 5) than the equivalent strain amplitude of LIGO  system noise in recent science runs; Figure 1  of \cite{Abbott:2006zx} shows strain amplitude spectra as low as $h_{rms}\simeq 5\times 10^{-23}/\sqrt{\rm Hz}$ near $\simeq 100$Hz.
The level of currently allowed excess strain noise from stochastic gravitational wave backgrounds is even lower, by about an order of magnitude.\cite{Abbott:2006zx} 

\begin{figure} 
\epsfysize=2.5in 
\epsfbox{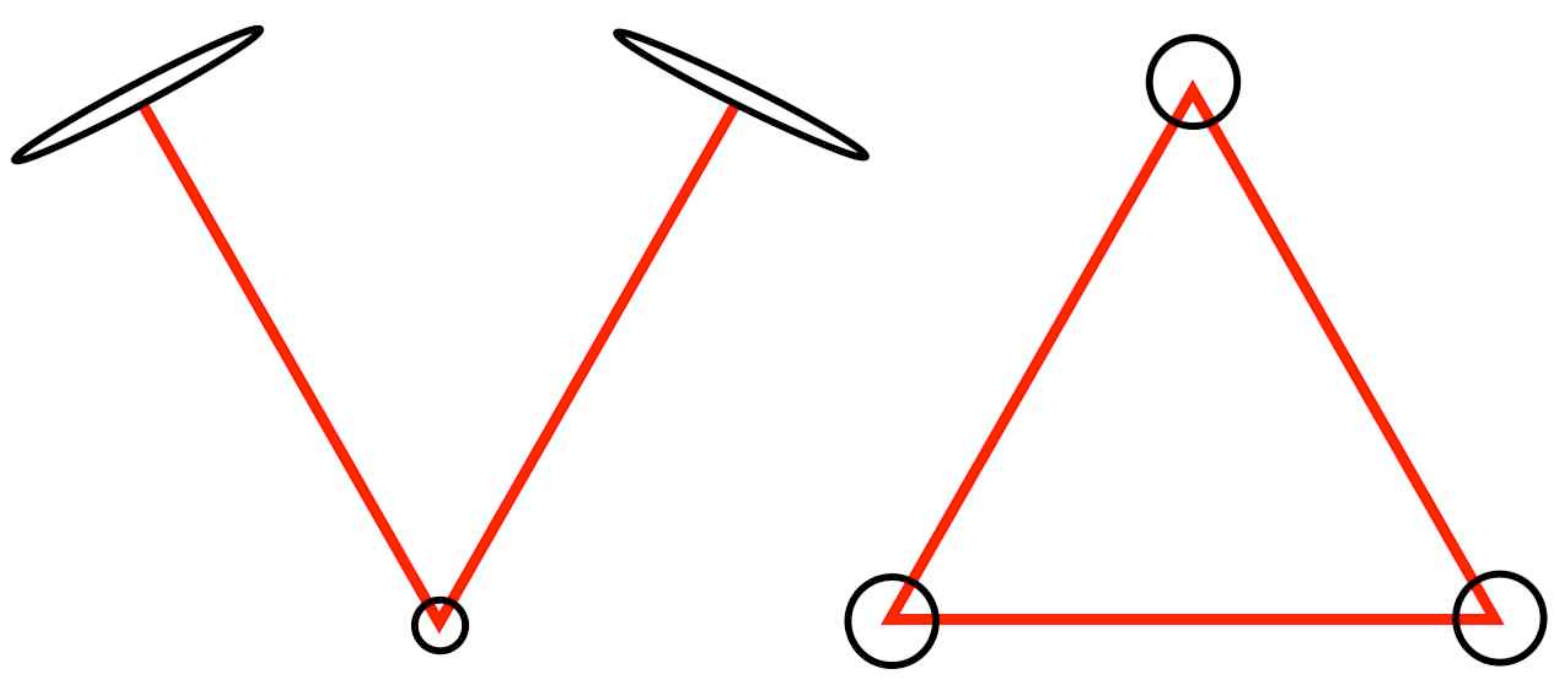} 
\caption{\label{squeeze} Sketch of the spacetime wavefunction shape at proof mass positions in  different interferometer configurations. At each vertex contours are shown of probability amplitude of transverse position in a particular spacetime state. The size of the contours  is greatly exaggerated, but the distortion is not.  A connected equilateral triangle (LISA-like configuration at right) has a minimum overall indeterminacy when the wavefunction is approximately isotropic at each vertex, leading to a lower bound on the level of holographic noise.  A Michelson setup (at left), with only two arms, can squeeze uncertainty into the unobserved direction, thereby minimizing the holographic indeterminacy (and noise) in the signal at the detected vertex. }
\end{figure}

At first glance this result seems to rule out the possibility of  holographic noise.
 However, holographic noise does not have the same effect as gravitational wave or instrument noise. In particular,  transverse positions of the distant proof masses are not actually measured in LIGO since it operates in a Michelson configuration,  so one Michelson  setup on its own is not sensitive to holographic noise.    The proof mass suspensions in LIGO   reduce radial noise as much as possible, thus they isolate degrees of freedom so that transverse noise does not leak into the signal. In operation the  system is placed into a state where holographic  indeterminacy  is    ``squeezed'' into the undetected transverse positions of the distant proof masses (see Fig. \ref{squeeze}).  It is possible that a small fraction  of transverse holographic noise leaks into the LIGO signal, perhaps even enough to contribute a detectable source of noise, but the system is not optimized to measure it.

\begin{figure} 
\epsfysize=2.5in 
\epsfbox{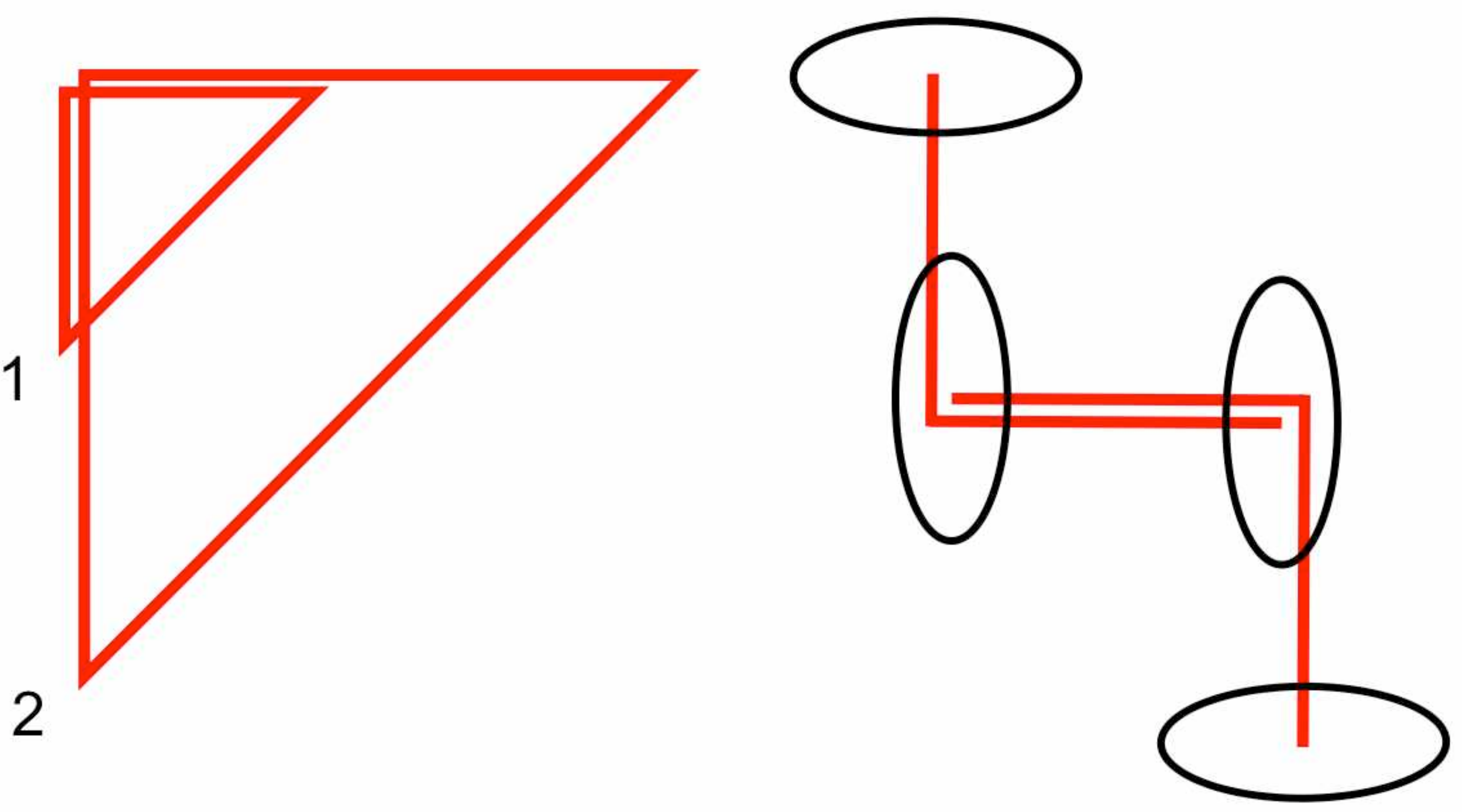} 
\caption{\label{detector}Sketch of   interferometer layouts  to detect holographic indeterminacy.  Transverse and radial positions  are simultaneously  measured in triangles (left) on different scales, 1 and 2 say, allowing studies of the amplitude and coherence of the holographic noise.  Alternatively (right),  a pair of Michelson interferometers can be arranged so as to avoid squeezing uncertainty out of the detected signals, since proof masses in close proximity share nearly the same spacetime wavefunction (whose contour shapes  are again sketched). Metrology with current techniques is adequate to detect the effect for baselines as small as $\simeq 1$m.}
\end{figure}

 On the other hand the radial positions are already measured more precisely than the holographic noise level,  and a  different configuration designed to be sensitive to transverse position  should detect the holographic noise. 
 A different interferometer design including   remote measurements over three arms, possibly   with shorter arms but with comparable measurement  precision in proof mass position, could detect the transverse holographic noise component.   The spacetime wavefunction shape is coherent over large  regions so 
 it is not necessary for all three arms to be part of the same phase-coherent interferometer.  That is, because positions commute to higher precision in smaller regions, proof masses in close proximity share similar spacetime wavefunctions: if one mass is in a squeezed state, a nearby mass in a separate interferometer shares that squeezing and its signal can detect that transverse uncertainty.  Thus two disconnected Michelson interferometers yield about the same result as one connected system  over similar paths, displaying  the quantum weirdness of holographic indeterminacy  (see Fig. \ref{detector}). Such a system might be optimized to operate at much higher frequencies than LIGO, on a scale of $\simeq 1$m or less, avoiding many sources of environmental noise. The flat spectrum predicted in Eq. (\ref{noise}) provides a clear observational target for design studies over a range of frequencies and coherence scales.
 
 Holographic noise  displays  quantum weirdness in a new way:  the spacetime wave function branching created by an interferometric measurement creates a macroscopically coherent difference, like   Schr\"odinger's cat, only applied to position instead of vitality. The difference itself is still microscopic of course--- the scale $\Delta x$ is much smaller than a Bohr radius--- but remarkably, applies coherently to the wavefunction of a macroscopic spacetime region.
Indeed, the positional displacement is coherent over a region of spacetime size $\simeq L$; the holographic noise in any two detectors in the same spacetime region should be highly correlated even if there is no connection between them other than spatiotemporal proximity.     The holographic effect could not be measured in any local experiment, but requires a long baseline separation to appear.   This  weirdness is imposed by the holographic lack of degrees of freedom.
  
\subsection{Measurement with LISA}

The proposed Laser Interferometer Space Antenna (LISA\cite{LISA,LISAconf}), with arm lengths $L\simeq 5\times 10^6$ km,  is an interferometer designed with a triangle configuration and therefore capable of making transverse position measurements.

\begin{figure} 
\epsfysize=3.5in 
\epsfbox{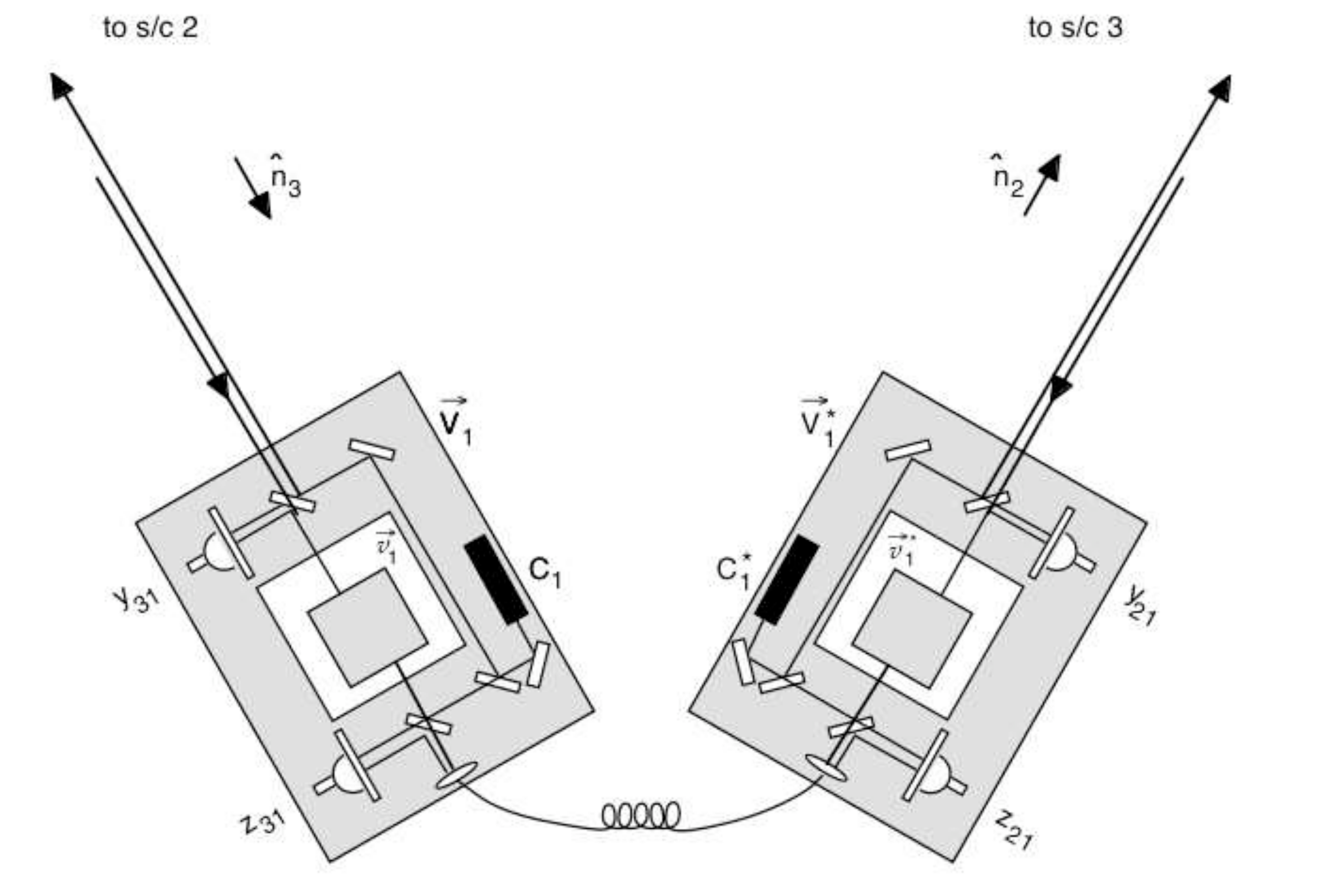} 
\caption{\label{schematicLISA}
 Schematic layout of one design for LISA proof masses and interferometers within each spacecraft, from\cite{LISA}. The scale of the spacecraft is $\simeq 1$ m and the arm length to the other spacecraft $\simeq 5\times 10^9$m. The  components are not drawn to scale and the detailed optical paths are only indicative.  In the text, these two masses in spacecraft 1 are named 12 and 13 according to their corresponding distant spacecraft. The interferometer measures the position of each proof mass only along its own radial direction ($\bf \hat n$). Along this axis, the position is measured   to the distant spacecraft.  For classical spacetime, this interferometer geometry produces an interferometer signal insensitive to transverse noise in the proof mass positions. However, the radial motion of one mass is partly along the transverse direction of the other, so if the two masses and their optical benches share the same (locally classical) coherent position state of spacetime metric, the transverse interdeterminacy relative to the distant partner of mass 12 shows up in the radial position of mass 13, and {\it vice versa}.}
\end{figure}

  Radial distances are measured between proof masses in each of LISA's three spacecraft across the three $\simeq 5\times 10^6$ km baselines.   
The detailed configuration  of LISA proof masses and optical benches within each spacecraft is not yet finalized but it may be  that there will be a separate proof mass for each beam, two within each spacecraft
(see Fig. \ref{schematicLISA}). 
  For holographic uncertainty, the situation reduces in any case to the same as a single mass per spacecraft  since the two proof masses in close proximity  share  the same local  classical spacetime state, so that the transverse uncertainty of one shows up  in the measured radial displacement signal of its partner, whether or not the transverse degrees of freedom of the masses themselves are precisely monitored.   With the interpretation here, this must be  the case since  position observables all commute locally.  With all three arms working, holographic indeterminacy must then show up as a new  contribution to the  signal that we seek to measure,  the holographic  noise.  This signal is a nonlocal propagation effect, like a gravitational wave, although as seen below the signatures are different.
  
Let  $\Delta x$ and $\Delta r$ denote the standard deviation in the measured position  of each mass along its own transverse and radial directions, relative to a classical value that vanishes in the absence of holographic noise.  In spacecraft 1,   let 12 and 13 denote the 
two masses, labled by their corresponding distant spacecraft 2 and 3.    According to Eq. (\ref{uncertainty}), each of the  proof masses has a conjugate  transverse time-delayed positional uncertainty as viewed from  its distant partner,
\begin{equation}
\Delta x_{12}\Delta x_{21} = (l_PL)/2.
\end{equation}
In normal LISA operating mode the two masses in each spacecraft at a given time share almost the same classical metric, so if the spacetime wavefunction is isotropic we can also set 
\begin{equation}
\Delta r_{13}= \cos(30^\circ) \Delta x_{12}.
\end{equation}
Simultaneous local position measurements in the same spacecraft must (nearly) commute with each other in the usual way (whether or not they are actually measured by the system), so once a radial position local spacetime state is determined for mass 13 it also fixes the transverse position local spacetime state for mass 12 and vice versa.  This introduces observable variations  $\Delta r_{13}\simeq\cos(30^\circ)\Delta x_{12}$ and so on for the other pairs. Since the interferometer measures the changes in  $r$ between three pairs of masses, the transverse indeterminacy enters into the signal stream.  

Let $L_1$ denote the measured length of the arm opposite spacecraft 1 from the signals in spacecraft 2 and 3. Independent measurements (meaning, those separated by more than a light travel time) have a standard deviation $\Delta L_1$
from adding  in quadrature,
\begin{equation}
\Delta L_1^2=\Delta r_{23}^2+\Delta r_{32}^2= \cos^2(30^\circ)(\Delta x_{21}^2+\Delta x_{31}^2),
\end{equation}
 and likewise for the other arms. If all three arms are being measured, so all the pairs are uncertain, 
 \begin{equation}
\Delta L_1^2+ \Delta L_2^2+\Delta L_3^2=\cos^2(30^\circ)(\Delta x_{21}^2+\Delta x_{31}^2+\Delta x_{12}^2+\Delta x_{32}^2+\Delta x_{13}^2+\Delta x_{23}^2).
\end{equation}
The sum now includes all three conjugate pairs. For each conjugate pair the sum has a lower bound from Eq. (\ref{uncertainty}), so there is a lower bound on the summed uncertainty in the measured arms. 

This argument only works if both masses  in each spacecraft have $r$ measured relative to a distant spacecraft, and those spacecraft have $r$ measured relative to each other; this is what allows the transverse component to be measured.  If only one of the two arms linked to a given spacecraft  is working, the whole spacecraft is placed in a position eigenstate along the direction of working arm and the uncertainty does not appear since its transverse position is not measured. By the same token, if any one of the three arms is not working,  the holographic noise need not appear in the other two arms, because, like LIGO, the spacecraft can be placed in  local spacetime states corresponding to small uncertainty in the radial directions of the measured arms (see Fig. \ref{squeeze}).   The third arm    collapses the wavefunction of each spacecraft into a state where the uncertainty is unavoidably detected.

 In the case of LISA the coherence time is the light travel time across an $L\simeq 5$ million kilometer LISA arm, about 17 seconds (or 60 millihertz frequency).  The most sensitive frequency for LISA for gravitational waves is about  ten millihertz so LISA is operating in the low-frequency, infrequent sampling regime. Setting aside factors of the order of unity, the  holographic noise  signal has an amplitude spectrum given by Eq. (\ref{noise})
up to the high frequency turnover at $f\simeq 60$ mHz due to LISA arm length.   This should be compared with instrumental and environmental noise sources. Above about  1 millihertz, with a long LISA data stream (of order a year say),  the astrophysical gravitational-wave background from white dwarf binaries\cite{Farmer:2003pa} should not be confusion limited so the instrument noise dominates at relevant frequencies.  The  reference design shot noise level is about\cite{LISA,prince2002}  
$h_{rms} \simeq 4\times 10^{-21} /\sqrt{\rm Hz}$ (in  position units,
$\Delta x_{rms} \simeq 2\times 10^{-9}{\rm cm/\sqrt{Hz}}$).  

Thus, using conservative estimates, the universal holographic noise is less than raw instrument noise in LISA by about a factor of 20. With  many sampling times it is nevertheless   possible to estimate the instrumental noise sources well enough to detect the presence of the smaller extra holographic noise signal.  If the noise sources can be separately estimated\cite{Hogan:2001jn}, after an observing time $\tau$ it is possible to detect the holographic noise if   $h_H$ is  less than the instrument sources  by a factor of up to $\simeq (f\tau/2)^{1/4}$, or about a factor of 40 after a few years.  For the nominal estimate in Eq. (\ref{noise}) that amounts to about a 2$\sigma$ detection.

Detailed study of the effect requires  separation of holographic noise from other noise sources.  A full analysis requires additional modeling and careful experimental design, but it is clear that a number of diagnostic options exist. For example, the fact   that the  interferometer response from holographic indeterminacy is not the same as that produced by gravitational waves  can be seen simply from the fact that gravitational wave signals appear in a Michelson setup (with only two arms working) and holographic signals do not. (With two arms, it becomes a LIGO-like configuration where no measurement is made of one arm and the transverse degree of freedom is isolated from the measured one, allowing the spacetime states of the spacecraft to spread  in the unmeasured direction.)
Even with all three arms working there are distinguishable  differences in the behavior of different noise sources: at low frequencies, gravitational waves, which are area-preserving transverse modes, are invisible in a symmetric Sagnac signal combination that measures the difference in total pathlength of beams travelling in the two opposite directions all the way around the triangle\cite{armstrong,tintoarmstrong,estabrook,tinto,Hogan:2001jn,Cornish:2001bb,Tinto:2004yw,TDI}.  Holographic uncertainty still  appears as noise in the Sagnac signal since    signals  combined in one spacecraft are monitoring the transverse length of the opposite arm including its indeterminacy.
Most sources of instrument noise  appear in  both Michelson and Sagnac combinations, again differing from gravitational wave and holographic signatures.
The experimental  signature of the holographic noise is thus significantly different from other sources of noise or   stochastic gravitational-wave backgrounds.

It seems likely that  holographic noise is detectable with interferometers  if holographic entropy bounds apply in nature and manifest themselves according to the interpretation presented here.
The  character of the noise including  its spacetime correlations will reveal concrete signatures of the holographic  nature of spacetime quantization.  Detection of the effect will lead to a natural interpretation of holographic entropy bounds in terms of a classical spacetime metric emerging from Planck-scale waves, a construction that may allow a concrete contact of experiments with fundamental theories of quantum gravity.  Further study is warranted of new experiments capable of measuring holographic noise, including more detailed modeling  of the LISA response.

\acknowledgements
I am grateful to P. Bender for useful comments on details of LISA system design.

\end{document}